\documentclass[epjCONF]{svjour}
\usepackage{graphics}
\usepackage[varg]{txfonts} 
\usepackage[latin1]{inputenc}
\session-title{MESON2012 - 12th International Workshop on Meson Production, Properties and Interaction} 
\begin{document}
\title{Spectroscopy of \boldmath$\eta'$-nucleus bound states at GSI-SIS}
\author{Hiroyuki~Fujioka \inst{1,}\thanks{\email{fujioka@scphys.kyoto-u.ac.jp}} \and 
Kenta~Itahashi \inst{2} \and
Hans~Geissel\inst{3} \and
Ryugo~S.~Hayano \inst{4} \and
Satoru~Hirenzaki\inst{5} \and
Satoshi~Itoh \inst{4} \and
Daisuke~Jido \inst{6} \and
Volker~Metag \inst{7} \and
Hideko~Nagahiro \inst{5} \and
Mariana~Nanova \inst{7} \and
Takahiro~Nishi \inst{4} \and
Kota~Okochi \inst{4} \and
Haruhiko~Outa \inst{2} \and
Ken~Suzuki \inst{8} \and
Takatoshi~Suzuki\inst{4} \and
Yoshiki~K.~Tanaka \inst{4} \and
Helmut~Weick \inst{3}}

\institute{Department of Physics, Kyoto University, Kyoto, 606-8502, Japan
\and Nishina Center for Accelerator-Based Science, RIKEN, Saitama, 351-0198, Japan
\and GSI Helmholtzzentrum f\"{u}r Schwerionenforschung GmbH, D-64291 Darmstadt, Germany
\and Department of Physics, The University of Tokyo, Tokyo, 113-0033, Japan
\and Department of Physics, Nara Women's University, Nara 630-8506, Japan
\and Yukawa Institute for Theoretical Physics, Kyoto University, Kyoto, 606-8502, Japan
\and II. Physikalisches Institut, Universit\"{a}t Gie{\ss}en, D-35392 Gie{\ss}en, Germany
\and Stefan Meyer Institut f\"ur subatomare Physik, 1090 Vienna, Austria
}

\abstract{
The $\eta'$ meson mass may be reduced due to partial restoration of chiral symmetry.
If this is the case, an $\eta'$-nucleus system may form a nuclear bound state.
We plan to carry out a missing-mass spectroscopy experiment
with the $^{12}\mathrm{C}$($p$,$d$)
reaction at GSI-SIS. Peak structures corresponding to such a bound state may be observed even in an inclusive measurement, if the decay width is narrow enough.
} 
\maketitle
\section{Introduction}
The $U_A(1)$ anomaly in QCD is considered to explain the peculiarly large mass of the $\eta'$ meson (about $958\ \mathrm{MeV}/c^2$) among the pseudoscalar meson nonet.
It is known that the anomaly has an effect on the $\eta^\prime$ mass only through the spontaneous and/or explicit SU(3) chiral symmetry breaking~\cite{Jido,Lee}. 
Recently it has been suggested, if chiral symmetry is partially restored in nuclear matter, 
that the $\eta^\prime$ mass may be reduced in nuclei with a suppression 
of the magnitude of the quark condensate~\cite{Jido}.
According to Nambu--Jona-Lasinio model calculations~\cite{Costa,Nagahiro}, the mass reduction at normal nuclear density amounts to around $150\ \mathrm{MeV}/c^2$. The significant in-medium mass reduction can be regarded as a strong attraction between the $\eta'$ meson and the nucleus. Then, an $\eta'$-nucleus system may exist as a bound state~\cite{Nagahiro,Nagahiro12}.

It should be mentioned that the scattering length of the $\eta'N$ interaction is evaluated to be of the order of $0.1\ \mathrm{fm}$ from the measurement of the near-threshold $pp\to pp\eta'$ cross section at COSY-11~\cite{Moskal}, which means that the $\eta'N$ interaction is very weak.
While it seems that these two scenarios contradict each other,
a reaction spectroscopy of $\eta'$-nucleus bound states will put a constraint on the strength of 
the $\eta'$-nucleus interaction~\cite{Nagahiro12}.

As for the decay width, which is also an important property in discussing the feasibility of an experimental observation, a recent measurement on $\eta'$ photoproduction from nuclei by the CBELSA/TAPS Collaboration revealed the absorption width of the $\eta'$ meson~\cite{Nanova}. Through the $A$-dependence of the transparency ratio, defined as the ratio of the cross section of $\gamma A\to \eta' A'$ and that of $\gamma N\to \eta' N'$, the absorption width is found to be 15--25$\ \mathrm{MeV}$ at normal nuclear density for an average $\eta'$ momentum of $1050\ \mathrm{MeV}/c$.
Therefore, one can expect the decay width of $\eta'$ bound states in nuclei could be small as well.

Motivated by these latest theoretical and experimental improvements on the understanding of
the $\eta'$-nucleus interaction, we proposed a missing-mass spectroscopy experiment 
of $\eta'$-nucleus bound states at GSI in 2011~\cite{ItahashiLOI,ItahashiPTP}.

\begin{figure}
\begin{center}
\resizebox{0.75\columnwidth}{!}{%
\includegraphics{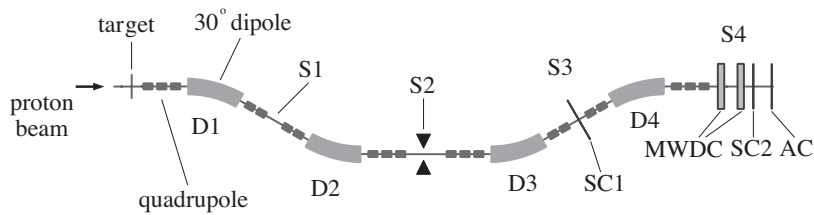} }
\caption{Schematic view of the FRS. See the text for the detail.}
\label{fig:1}       
\end{center}
\end{figure}
\section{Experimental Principle}
We will make use of the $^{12}\mathrm{C}$($p$,$d$) reaction for a missing-mass spectroscopy of $\eta'$-nucleus bound states. The incident proton beam with kinetic energy of $2.50\ \mathrm{GeV}$ will be 
supplied by the SIS synchrotron at GSI, and the ejectile deuteron will be momentum-analyzed by the fragment separator FRS (Fig.~\ref{fig:1}). With this experimental condition, a high-resolution and high-statistics spectroscopy will be enabled.

Another feature of the proposed experiment is to perform an inclusive measurement; we will detect only the outgoing deuteron in the ($p$,$d$) reaction, and not the decay particle of $\eta'$-mesic nuclei. Then, an unbiased spectrum can be obtained without assuming properties of the decay process despite a poorer signal-to-noise ratio. As shown later, a high sensitivity to observe a peak structure over a huge physical background is expected only if the decay width of $\eta'$-mesic nuclei is narrow enough. At present, this inclusive measurement for $\eta'$-nucleus bound states is considered only at GSI-SIS.

We will detect deuterons at the final focal plane S4 of FRS, as the FRS optics will be set to momentum-dispersive. Multi-wire drift chambers (MWDC) installed at S4 will serve for momentum measurement. For the particle identification, an Aerogel \v{C}erenkov counter (AC) with a high refractive index around 1.20~\cite{Adachi} will be installed as a veto counter for the predominant background of protons from the ($p$,$p'$) reaction on the target.
Furthermore, time-of-flight between a set of segmented plastic scintillators (SC1 at S3 and SC2 at S4) will be used in the offline analysis. Another source of background, which consists of secondary particles from the beam pipe near S1 due to the intense proton beam, needs to be removed between S1 and S3, in order to decrease the particle rate at S4. One of the possible solutions for this is to set the optics of the first half of FRS (S0--S2) as the momentum compaction mode and to install a slit at S2.

The overall spectral resolution is estimated to be $\sigma=1.6\ \mathrm{MeV}$, which is sufficiently smaller than the expected width of $\eta'$-mesic nuclei. Such a high resolution is an indispensable requirement in realizing an inclusive measurement.

\section{Expected Spectrum}
\begin{figure}
\begin{center}
\resizebox{\columnwidth}{!}{%
\includegraphics{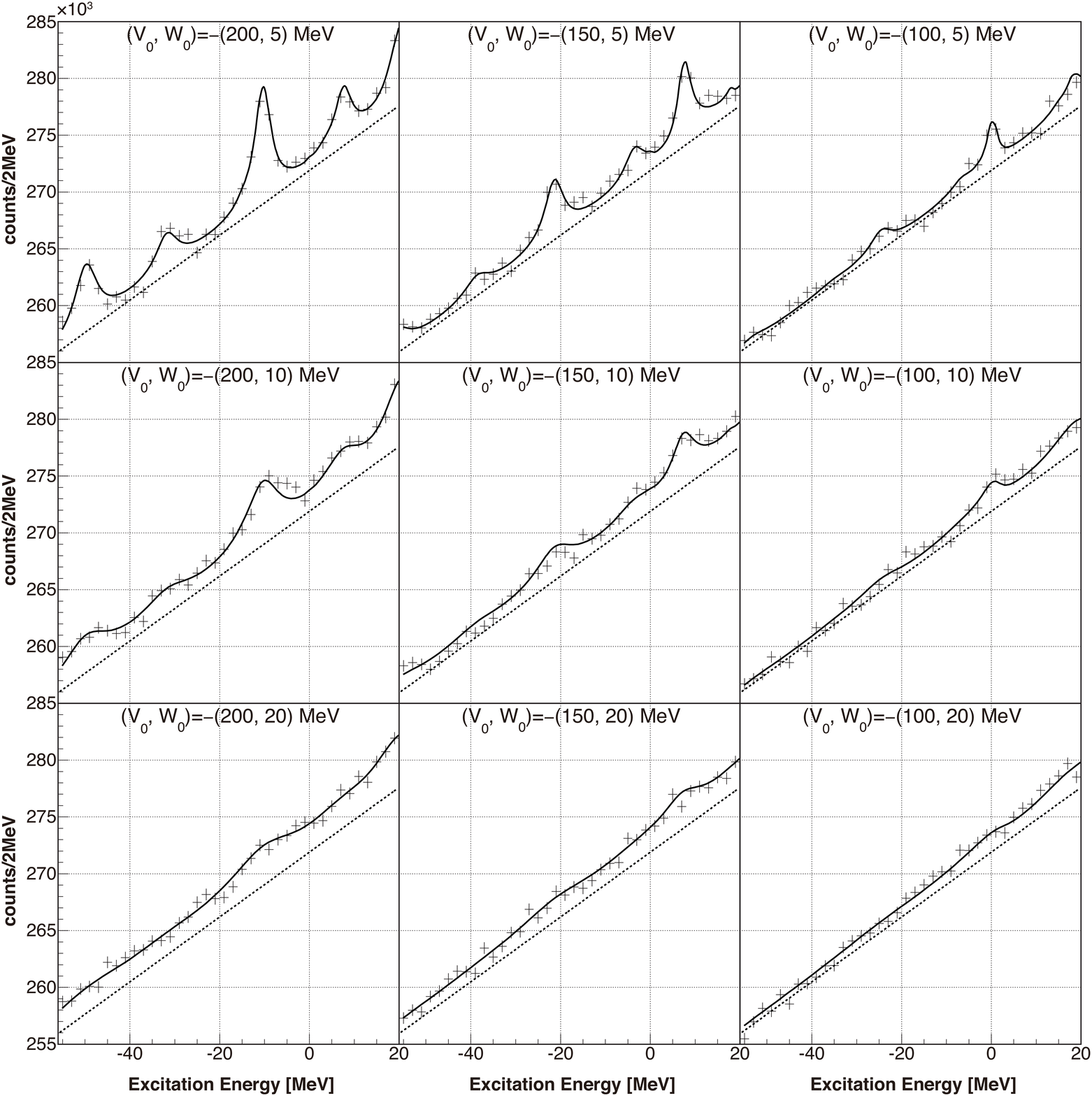} }
\caption{Simulated spectra for different $\eta'$-nucleus optical potential ($V_0+iW_0$). The dashed line corresponds to the background quasi-free processes.}
\label{fig:2}       
\end{center}
\end{figure}

The formation cross section of the $p+{}^{12}\mathrm{C}\to d+{}^{11}\mathrm{C}\otimes \eta'$ is calculated by the Green's function method~\cite{Hirenzaki},
in which $\eta'$-nucleus optical potential $V_{\eta'}$ is assumed to be proportional to the nuclear density $\rho(r)$, that is, $V_{\eta'}=(V_0+iW_0)\rho(r)/\rho_0$, where $\rho_0$ is the normal nuclear density. Although there is no experimental data available on the cross section of the elementary process $pn\to d\eta'$, we have evaluated it to be around $3\ \mathrm{\mu b}$ in the following two ways.
\begin{enumerate}
\item
We assume
$\sigma(pn\to d\eta')/\sigma(pp\to pp\eta')=\sigma(pn\to d\eta)/\sigma(pp\to pp\eta)$,
according to the prescription in Ref.~\cite{Rejdych}. The other three processes have been measured by 
CELSIUS/WASA~\cite{Calen} and COSY-11~\cite{Khoukaz}.
\item
Grishina \textit{et al.}~\cite{Grishina} calculated the cross section with a two-step model, taking into account not only $\pi$ but also $\rho$, $\omega$ exchange. The result is $\sim 3\ \mathrm{\mu b}$ with an uncertainty of factor 1.5.
\end{enumerate}
Assuming an isotropic distribution in the $pn\to d\eta'$ reaction, the forward differential cross section in the laboratory frame is obtained as $\sim 30\ \mathrm{\mu b/sr}$ for incident kinetic energy of $2.5\ \mathrm{GeV}$. The detail of the theoretical calculation with the Green's function method will be given in Ref.~\cite{Hirenzaki}.

The background in the inclusive ($p$,$d$) spectrum mainly comes from the quasi-free process such as multi-pion production by the $pp\to dX$ or $pn\to dX$ reaction. The total background level is estimated to be around $4\ \mathrm{\mu b/sr/MeV}$ by using $\omega$ production data by COSY-ANKE~\cite{Barsov} as a reference.

Combining the signal and background, we carried out a simulation with $3.24\times 10^{14}$ protons on a $4\ \mathrm{g/cm^2}$-thick carbon target\footnote{Corresponds to the beamtime request in the Letter of Intent~\cite{ItahashiLOI}.}. The result for various types of optical potentials is shown in Fig.~\ref{fig:2}. The signal-to-noise ratio is found to be of the order of 1/100 at most. In case that the real part $|V_0|$ is large and the imaginary part $|W_0|$ is small, clear peak structures corresponding to $\eta'$-mesic nuclei can be observed, especially near the $\eta'$ production threshold. They could be a signature of an attractive $\eta'$-nucleus interaction.

It should be stressed that the Nambu--Jona-Lasinio calculations correspond to $|V_0|\sim 150\ \mathrm{MeV}$, and that the transparency ratio measurement by CBELSA/TAPS indicates $|W_0|\lesssim 12.5\ \mathrm{MeV}$. If so, we may be able to observe at least one peak structure due to an excited state of $\eta'$-mesic nuclei close to the threshold.

\section{Summary}
We plan to perform a spectroscopy experiment of $\eta'$-mesic nuclei with the $^{12}\mathrm{C}$($p$,$d$) reaction at GSI-SIS. A simulation based on a theoretical calculation and a background estimation demonstrated the experimental feasibility of an inclusive measurement. When the mass reduction of the $\eta'$ meson in medium is significant as the Nambu--Jona-Lasinio calculations suggest and the decay width of $\eta'$-nucleus bound states is small, we may be able to observe a peak structure near the $\eta'$ production threshold, which suggests a strong attraction between an $\eta'$ meson and a nucleus.

After submission of the Letter of Intent~\cite{ItahashiLOI} in 2011, we started to develop the aerogel \v{C}erenkov counter, the FRS optics optimized for this experiment, and so on. We expect the data acquisition can be done in 2013--2014.
\begin{acknowledgement}
This work is partly supported by the Grant-in-Aid for Scientific Research on Innovative Areas
(No. 24105705)
from the Ministry of Education, Culture, Sports, Science and Technology of Japan.
\end{acknowledgement}

\end{document}